\documentclass[preprint,aps,showpacs,nofootinbib,preprintnumbers,amsmath,amssymb]{revtex4-1}
\usepackage{amsfonts}
\usepackage{amssymb}
\usepackage{}
\usepackage{epsfig}
\usepackage{graphicx}% Include figure files
\usepackage{subfigure}
\usepackage{dcolumn}% Align table columns on decimal point
\usepackage{bm}% bold math
\usepackage[usenames ,dvipsnames]{xcolor}
\usepackage{slashed}

\begin{document}

\title{X-ray Line from the Dark Transition Electric Dipole}

\author{Chao-Qiang~Geng\footnote{geng@phys.nthu.edu.tw},
Da~Huang\footnote{dahuang@phys.nthu.edu.tw}
and
Lu-Hsing Tsai\footnote{lhtsai@phys.nthu.edu.tw}}
  \affiliation{
 %College of Science,
  Chongqing University of Posts \& Telecommunications, Chongqing, 400065, China\\
 Department of Physics, National Tsing Hua University, Hsinchu, Taiwan\\
 Physics Division, National Center for Theoretical Sciences, Hsinchu, Taiwan
}

\date{\today}

\begin{abstract}
We study a two-component dark matter (DM) model in which the two Majorana fermionic DM components with nearly degenerate masses are stabilized by an $Z_2$ symmetry and interact with the right-handed muon and tau only via real Yukawa couplings, together with an additional $Z_2$-odd singly-charged scalar. In this setup, the decay from the heavy DM to the lighter one via the transition electric dipole yields the 3.55 keV X-ray signal observed recently. The Yukawa couplings in the dark sector are assumed to be hierarchical, so that the observed DM relic abundance can be achieved with the leading s-wave amplitudes without a fine-tuning. We also consider the constraints from flavor physics, DM direct detections and collider searches, respectively.
\end{abstract}

%\pacs{95.35.+d, 98.70.Sa, 13.85.Tp, 14.80.-j}
%\keywords{}
\maketitle
\section{Introduction}\label{intro}
Dark Matter (DM), although there are overwhelming evidences of its existence from astrophysics and cosmology, is still mysterious from the particle physics point of view and is one of few indications of physics beyond the Standard Model (SM)~\cite{Bertone:2004pz}. Traditionally, one hopes to find some of its properties either by observing nuclear recoils in the underground direct detection experiments~\cite{Freese:2012xd,Jungman:1995df,Smith:1988kw} or by the indirect searches via the measurements of cosmic and gamma rays~\cite{Cirelli:2012tf}. The most recent possible signal comes from the discovery of the unidentified X-ray line at about 3.55 keV from the observation of the Andromeda galaxy and many galaxy clusters, including the Perseus galaxy cluster by XMM-Newton X-ray Space observatory~\cite{Bulbul:2014sua,Boyarsky:2014jta}, which have inspired particle physicists to propose many interesting DM candidates~\cite{Bulbul:2014sua,Boyarsky:2014jta,GeneralModel}.

In this paper, we use a simple two-component DM model~\cite{MultiComponent} to explain the 3.55 keV X-ray line. The dark sector contains two Majorana fermionic DM fields and a heavy singly-charged scalar which enables the DMs to couple to the right-handed $\mu$ and $\tau$ leptons. Both the DM and singly-charged scalar fields are charged under a $Z_2$ symmetry to stabilize the two DM particles. The decay of the heavier DM particle to its lighter sister can only proceed via the dark transition electric dipole moment (EDM), resulting in the observed X-ray signal and guaranteeing the stability of the heavy DM within the Universe age. Furthermore, because of the nearly degenerate DM masses and the particular choice of the Yukawa structure, the correct DM relic abundance is achieved via the thermal production by the coannihilation of the two DM particles into $\mu^\pm \tau^\mp$ pairs in the $s$-wave with the almost one half for each component.

The paper is organized as follows. In Sec.~\ref{model}, we show the relevant Lagrangian of the DM sector. We then study the X-ray line
 in Sec.~\ref{Xray} and the DM relic abundance in Sec.~\ref{DMRelic}. In Sec.~\ref{constraint}, 
 we consider constraints on the model from flavor physics, DM direct detections and collider searches. In Sec.~\ref{result}, we present our numerical results. Our conclusions and discussions are given in 
Sec.~\ref{conclusion}.

\section{The Model}\label{model}
In our model, in addition to the SM particle content, we introduce two Majorana fermionic DM fields $\chi_{1,2}$, which are neutral under the SM gauge groups, and a heavy singly-charged scalar mediator $\eta$ to allow the two DM fields to interact with the right-handed leptons. For simplicity, we restrict ourselves to the case where the two DMs only couple to the two heavier families of leptons, $\mu$ and $\tau$. Both $\chi_{1,2}$ and $\eta$ are odd under the dark $Z_2$ symmetry which stabilizes the DM particles. The quantum numbers of the relevant fields are summarized in Table~\ref{charge}.
\begin{table}\caption{Quantum Numbers for Relevant Fields}
\begin{center}
\begin{tabular}{c|c c c}\label{charge}
   & $SU(2)_L$ & $U(1)_Y$ & $Z_2$ \\
  \hline
  $\ell_R$ & ${\mathbf 1}$ & $-2$ & + \\
  $\chi_{1,2}$ & ${\mathbf 1}$ & 0 & $-$ \\
  $\eta$ & ${\mathbf 1}$ & $-2$ & $-$ \\
\end{tabular}
\end{center}
\end{table}
With the assigned quantum numbers for the new particles, we can write down the general renormalizable Lagrangian related to the dark sector as follows:
\begin{eqnarray}
{\cal L}_{\rm DM} &=& \sum_{i=1,2}\bar{\chi}_i \slashed{\partial} \chi_i - \frac{1}{2}\sum_{i,j=1,2} \bar{\chi}^c_i M_{ij} \chi_j + (D^\mu \eta)^\dagger (D_\mu \eta)-V(\eta) - \sum_{\substack{i=1,2\\ \ell=\mu,\tau}}(h_{\ell i}\bar{\ell}P_L \chi_i\eta +{\rm h.c.}),\nonumber\\
V(\eta) &=& m_\eta^2\,\eta^\dagger\eta+ \frac{\lambda}{2}(\eta^\dagger\eta)^2,\, \mbox{and}\,\, D_\mu \eta = (\partial_\mu - i e A_\mu)\eta ,
\end{eqnarray}
where $M$ denotes the Majorana mass matrix for the two DM fields, $P_L=({\mathbb I}-\gamma^5)/2$ the usual left-handed projection operator and $A_\mu$ the electromagnetic field. We further assume that the Yukawa couplings $h_{\ell i}$ are all real in the bases of the DM and lepton mass states. %{\color{red} 
Furthermore, the constraints from the lepton flavor violating process $\tau\to\mu\gamma$ and the DM relic density force us to take the hierarchical form of the Yukawa couplings $h_{\mu 1} \sim h_{\tau_2} \gg h_{\mu 2} \sim h_{\tau 1}$. For simplicity, we assume the equal value $h_L(h_S)$ of the (off-)diagonal entry: 
%}
\begin{equation}\label{YukMat}
h_{\ell i} = \left(\begin{array}{cc}
                    h_L & h_S \\
                    h_S & h_L
                  \end{array}
\right),\;\;h_L\gg h_S\;.
\end{equation}
{\color{red}  }

\section{X-Ray Line}\label{Xray}
It is well-known that the dipole operator of a Majorana particle vanishes. However, for a model with multiple neutral Majorana particles there still exist the transition dipole moments. In this section, we will show that the decay of the heavy Majorana DM particle into the lighter one via the transition electric dipole operator can naturally explain the cluster X-ray line at 3.55 keV. We start with the effective dipole operator in our model
\begin{eqnarray}\label{dipole}
{\cal O}_{\rm dipole} = \frac{1}{2}\bar{\chi}_1\sigma^{\mu\nu}[D_\chi\gamma^5+i\mu_\chi\mathbb{I}]\chi_2 F_{\mu\nu},
\end{eqnarray}
where $D_\chi$ and $\mu_\chi$ denote the dark transition electric dipole moment (EDM) and transition magnetic dipole moment (MDM), respectively.
\begin{figure}[ht]
\centering
\includegraphics[width=0.47\textwidth]{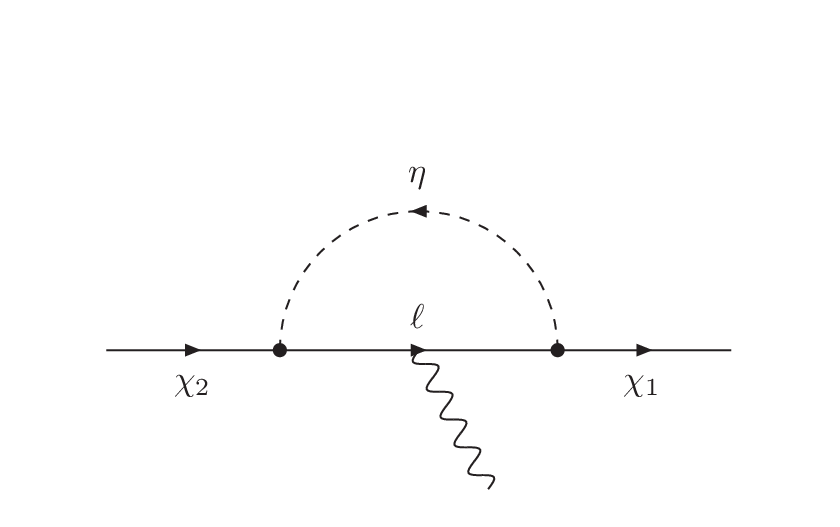}
\includegraphics[width=0.47\textwidth]{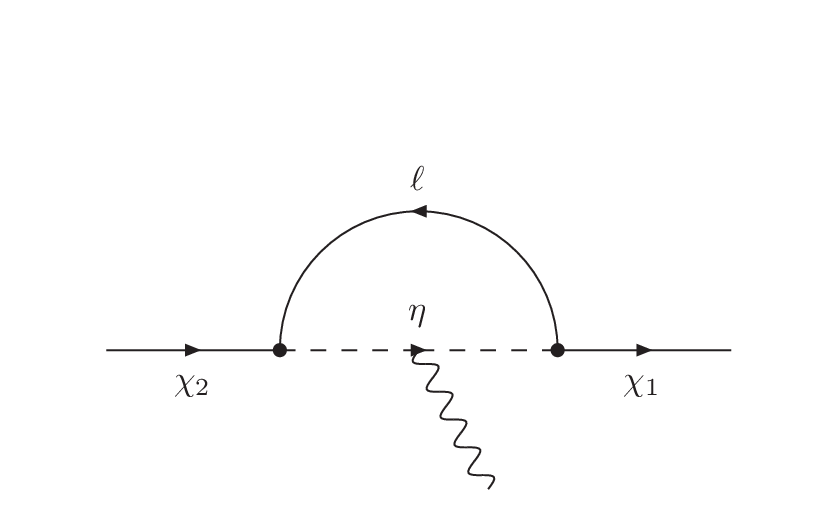}
\caption{Feynman diagrams for the transition dipole operator between $\chi_{2,1}$.}
\label{Fig_Dip}
\end{figure}
From Fig.~\ref{Fig_Dip}, we obtain
\begin{eqnarray}\label{EDM}
D_\chi = \frac{e R}{16\pi^2} \frac{\Delta m}{2 m_\chi^2} \Big[1+ \frac{m_\eta^2}{m_\chi^2}\ln\big(1-\frac{m_\chi^2}{m_\eta^2} \big) \Big] \approx \frac{e R }{64\pi^2} \frac{\Delta m}{ m_\eta^2}, \quad \mu_\chi =0,
\end{eqnarray}
where $m_\chi$ ($\Delta m$) stands for the average mass (the mass difference) of the two DM particles and $R=h_{\mu 1}h_{\mu 2}+ h_{\tau 1} h_{\tau 2} = 2 h_L h_S$. The decay rate of the heavier DM is then given by
\begin{equation}\label{DecRate}
\Gamma = \frac{\Delta m^3}{\pi} D_\chi^2 \approx \frac{\Delta m^3}{\pi} \Big( \frac{eR}{64\pi^2} \frac{\Delta m}{m_\eta^2} \Big)^2.
\end{equation}
Note that the vanishing value of the transition MDM can be traced back to our assumption of the real Yukawa couplings. However, if this assumption is relaxed, then the MDM would have a nonzero value of $\mu_\chi$, given by
\begin{eqnarray}
\mu_\chi = \frac{ e I}{16\pi^2} \frac{1}{m_\chi} \left[1+ \frac{m_\eta^2}{m_\chi^2}\ln\big(1-\frac{m_\chi^2}{m_\eta^2} \big) \right] \approx \frac{e I }{32\pi^2} \frac{m_\chi}{m_\eta^2},
\end{eqnarray}
where $I= {\rm Im} (h^*_{\mu 1}h_{\mu 2}+ h^*_{\tau 1} h_{\tau 2})$. Clearly, if the imaginary part of the Yukawa couplings is of the same size as the real one, the MDM would be the main contribution since $\mu_\chi/D_\chi \propto \Delta m/m_\chi \sim {\cal O}(10^{-7})$.

In order to make numerical studies, we need to transform the lifetime presented in Ref.~\cite{Boyarsky:2014jta} into our setup. The prototype model in Refs.~\cite{Boyarsky:2014jta,Bulbul:2014sua} contains the sterile neutrino with the mass $m_0=7.1$ keV, which decays into an active neutrino plus a photon via the small transition magnetic moment with the lifetime in the range $\tau_0 = 0.2 \sim 2\times 10^{28}~{\rm s}$. In our present model, we rescale the above lifetime to match the observed X-ray flux to be
\begin{eqnarray}\label{lifetime}
\tau_{\chi} = \frac{\tau_0 m_0}{2 m_\chi},
\end{eqnarray}
where the factor 2 accounts for the fact that each DM component carries only half of the DM density in the whole Universe and the Galaxy. From Eq.~(\ref{lifetime}), we see that the predicted lifetime depends on the average DM mass $m_\chi$. With Eqs.~(\ref{DecRate}) and (\ref{lifetime}), 
we will determine the parameter space in our model to explain the cluster X-ray line.
% as shown in the orange region bounded by two black curves in Fig.~\ref{Fig_Result} for several benchmark points 
%with the fixed ratios of $R_M=m_\eta/m_\chi$ and $R_h = h_S/h_L$.

%{\color{red} 
The measurement of the cluster X-ray line gives some constraint to the mass of the charged scalar $m_\eta$. 
 From Eqs.~(\ref{DecRate}) and (\ref{lifetime}), we can derive the following formula:
\begin{eqnarray}
m_\eta = \left[ \frac{\Delta m^5}{\pi} \left(\frac{eR}{64\pi^2}\right)^2 \left(\frac{\tau_0 m_0}{2 m_\chi} \right) \right]^{\frac{1}{4}},
\end{eqnarray}
If we fix the values of the dark matter mass and the coupling ratio, for example, $m_\chi=10$~GeV and $R_h=h_S/h_L=0.1$, 
it is interesting to see that there is an upper limit on  $m_\eta$ in the light of the cluster X-ray anomaly and the perturbativity of the Yukawa coupling. 
If we further take the largest lifetime $(\tau_0)_{\mathrm max}=2 \times 10^{28}$s and the maximal value $h_L=5$ allowed by the perturbativity, 
we obtain this upper limit to be $1.86$~TeV. Yet, this result comes solely from the consideration of the perturbativity and the measured X-ray flux. If we consider more constraints from other experiments, the upper bound on $m_\eta$ would further decreased, 
%which can be indicated from Fig.~\ref{Fig_Result}
which will be discussed in Sec.~V.
%}

\section{Dark Matter Relic Abundance}\label{DMRelic}
It is clear that a successful prediction of the correct DM relic abundance is one of the most important tests for the viability of a DM model. Usually, for a cold DM, the generation of its relic abundance is assumed to proceed by the thermal freezing-out mechanism, which relies on the thermally averaged effective DM annihilation cross section at the time of the freeze-out. Since the two DMs have the highly degenerate masses by the construction, the coannihilations between $\chi_1$ and $\chi_2$ have to be considered~\cite{Griest:1990kh}.

\begin{figure}[ht]
\centering
\includegraphics[width=0.45\textwidth]{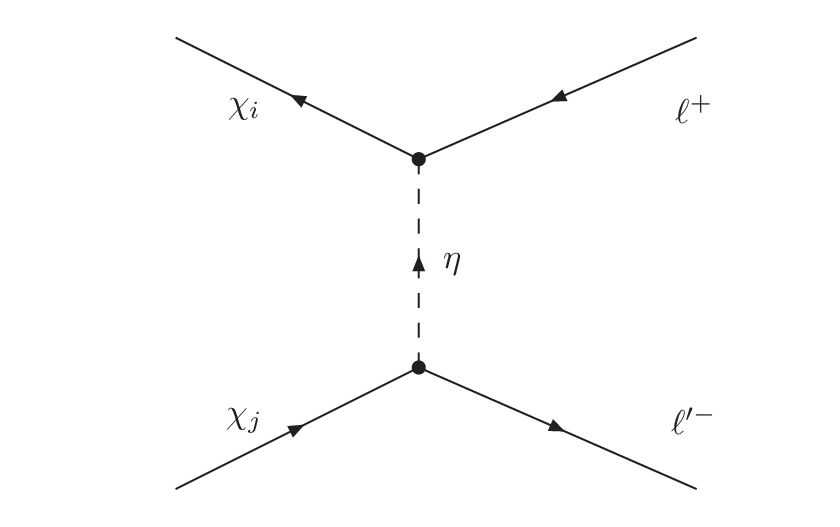}
\includegraphics[width=0.45\textwidth]{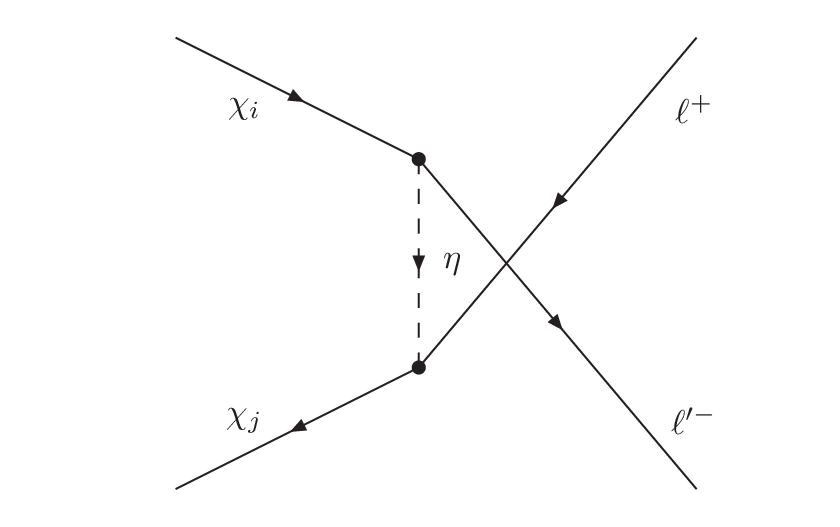}
\caption{Annihilation diagrams of the DM particles.}
\label{Fig_Ann}
\end{figure}

For a fermionic DM with its freezing-out mainly through $t/u$-channel annihilations, it is well-known that the thermally averaged effective annihilation is dominated by the $p$-wave contribution, which usually requires large couplings and/or light mediators to enhance the annihilation so as not to over-close the Universe. Let us make this point more precise by looking into the leading tree-level Feynman diagrams in Fig.~\ref{Fig_Ann}.
Since the annihilation with the definite incoming DMs and outgoing leptons involves $t$- and $u$-channels, if the Yukawa couplings in the two diagrams are the same, the destructive interference between the two channels would cause the leading $s$-wave contribution to be vanishingly small. However, in the presence of the two DM components with the Yukawa structure in Eq.~(\ref{YukMat}), there is a special process, $\chi_1\chi_2 \rightarrow \mu^\pm \tau^\mp$, which does not suffer such a suppression. The $t$-channel diagram is only proportional to $h_{\mu 1}h_{\tau 2}=h_L^2$, while the corresponding $u$-channel $h_{\tau 1}h_{\mu 2}=h_S^2$. Based on our construction of $h_L\gg h_S$, the $t$-channel can not be fully cancelled by the corresponding $u$-channel, leading to a large dominant $s$-wave effective cross section:
\begin{eqnarray}\label{swave}
\langle\sigma v_{\rm rel}\rangle = \sigma_0 = \frac{(4m_\chi^2-m_\tau^2)^2}{512\pi m_\chi^4 (2m_\chi^2+2m_\eta^2-m_\tau^2)^2} \left(h_L^4(4m_\chi^2+m_\tau^2)-8h_L^2 h_S^2 m_\chi^2+h_S^4(4m_\chi^2+m_\tau^2)\right).\nonumber\\
\end{eqnarray}
In the following numerical computation of the DM relic abundance, we shall use Eq.~(\ref{swave}) as our effective annihilation cross section. As a result, the analytic solution of the Boltzmann equation describing the evolution of the DM density is given by~\cite{Kolb94,ArkaniHamed:2006mb}:
\begin{equation}\label{RelicDens}
\Omega h^2 = \frac{688 \pi^{5/2} T^3_\gamma x_f }{99\sqrt{5g_\star}(H_0/h)^2 M^3_{\rm Pl} \sigma_0 } = \frac{8.7\times 10^{-11} {\rm GeV}^{-2} x_f }{\sqrt{g_\star} \sigma_0},
\end{equation}
where the freeze-out temperature $T_f$ is implicitly defined in $x_f=m_\chi/T_f$, given by
\begin{equation}
x_f = X-\frac{1}{2} \log X, \quad X=25+\log\Big[\frac{g}{\sqrt{g_\star(T_f)}} m_\chi \sigma_0 \times 6.4\times 10^6 {\rm GeV} \Big]\,,
\end{equation}
and $g_\star(T_f)$ denotes the number of relativistic degrees of freedom at $T_f$, taken from the tabulated functions in Refs.~\cite{Srednicki:1988ce,Olive:1980wz}. The final result is shown as the thick red curve in the $m_\chi$-$h_L$ plane for several selected benchmark mass and coupling ratios in Fig.~\ref{Fig_Result}, where we have used the central value of the most recent Planck result $\Omega_{\rm DM} h^2 = 0.1187\pm0.0017$~\cite{Planck}. Finally, we remark that the neglect of the other annihilation channels to the effective cross sections does not affect the final prediction of the DM relic abundance much, with the correction within 5\% at best, since they are $p$-wave suppressed as mentioned before.

\section{Complementary Constraints on the Model}\label{constraint}
In the previous sections, we have shown that our two-component DM model can not only explain the X-ray line observed from the distant galaxy clusters but also naturally give rise to the desired DM relic abundance in the Universe. However, it is clear that in order for this scenario to be viable, we have to investigate if it satisfies with the constraints from other aspects of particle physics, such as flavor physics, direct DM detections, and collider searches.

\subsection{Flavor Physics}
The most stringent flavor constraint on the model comes mainly from the lepton flavor violation (LFV) process $\tau\to \mu\gamma$, which proceeds via the dipole operator $\frac{1}{2} \bar{\mu} \sigma^{\mu\nu} (A_L P_L+A_RP_R)\tau F_{\mu\nu}$ with the coefficients $A_{L,R}$:
\begin{equation}
A_L = \frac{e}{16\pi^2}(2h_L h_S) F_2\Big(\frac{m_\chi^2}{m_\eta^2}\Big)\frac{m_\tau}{2m_\eta^2}, \quad
A_R = \frac{e}{16\pi^2}(2 h_L h_S) F_2\Big(\frac{m_\chi^2}{m_\eta^2}\Big)\frac{m_\mu}{2m_\eta^2},
\end{equation}
where the function $F_2(x)$ is given by
\begin{equation}
F_2(x) = \frac{1-6x+3x^2+2x^3-6x^2\log x}{6(1-x)^4}.
\end{equation}
The branching ratio for this LFV process is~\cite{Schmidt:2012yg}:
\begin{eqnarray}\label{TMG}
{\mathcal B}(\tau\to \mu\gamma) = \frac{3\alpha_{\rm em}}{64\pi G_F^2 m_\eta^4} \Big|(2 h_L h_S)F_2\left(\frac{m_\chi^2}{m_\eta^2}\right)\Big|^2 {\mathcal B}(\tau\to \mu\nu_\tau\bar{\nu}_\mu),
\end{eqnarray}
where $\alpha_{\rm em} = e^2/(4\pi)$ and $G_F$ are the fine structure and Fermi constants, respectively. In Eq.~(\ref{TMG}), we have neglected the terms proportional to the lepton masses.
Using the current experimental upper bound of ${\mathcal B}(\tau\to\mu\gamma)<4.4\times 10^{-8}$~\cite{PDG}, we present the constrained parameter space as the blue region with the dashed line as the boundary in Fig.~\ref{Fig_Result}. From the figure, we also see that a lot of parameter space is already excluded by this LFV channel, especially when the hierarchy between $h_S$ and $h_L$ is not very large, {\it i.e.}, $h_S/h_L \sim 0.2$.

\subsection{Direct Dark Matter Detections}
Currently, the most stringent cross section bound for the dark matter mass range of interest is from the LUX experiment~\cite{LUX}, in which for the 33 GeV dark matter the spin-independent cross section bound is of ${\cal O} (10^{-46})$~${\rm cm}^2$. For the present leptophilic DM model, the relevant scattering of two DM components with the target nucleus is proceeded by the mediation of a virtual photon. As discussed in the literature~\cite{Bai:2014osa,Agrawal:2011ze,Agrawal:2014ufa,Chang:2014tea,Fitzpatrick:2010br,Ho:2012bg,DelNobile:2014eta}, we first need to identify the relevant effective operators with the DMs coupled to a photon, and then calculate the scattering probability of the DM particles with the target nucleus. Each DM can be viewed as a Majorana particle in the mass eigenstate basis $\chi_i$, so that if the scattering does not change the DM component, for the so-called component-conserving scattering, one has to only consider the following two kinds of dimension-six effective operators
\begin{eqnarray}
{\cal O}^c_{1 i} = -\bar{\chi_i} \gamma^\mu \gamma^5 \partial^\nu \chi_i F_{\mu\nu}+ {\rm h.c.}, \quad\quad {\cal O}^c_{2 i} = i \bar{\chi_i}\gamma^\mu \partial^\nu \chi_i F^{\alpha\beta}\epsilon_{\mu\nu\alpha\beta} + {\rm h.c.},
\end{eqnarray}
where $i=1$ and 2. However, as pointed in Ref.~\cite{Bai:2014osa}, it can be proved that ${\cal O}^c_{2i} = -2 {\cal O}^c_{1i}$ by using the Chisholm identity, which results in a single kind of the dimension-six operator for the Majorana fermions. This kind of operators can be matched to the electromagnetic anapole moment of each DM component, which further couples to the current from the target nucleus in the non-relativistic limit and realizes the anapole dark matter~\cite{Bai:2014osa,Fitzpatrick:2010br,Ho:2012bg,DelNobile:2014eta}. However, besides the above component-conserving interactions, there are also transition interactions induced by the real Yukawa couplings,
\begin{eqnarray}
{\cal O}^t_1 = -\bar{\chi_2} \gamma^\mu \gamma^5 \partial^\nu \chi_1 F_{\mu\nu}+ {\rm h.c.}, \quad\quad {\cal O}^t_2 = i \bar{\chi_2}\gamma^\mu \partial^\nu \chi_1 F^{\alpha\beta}\epsilon_{\mu\nu\alpha\beta} + {\rm h.c.},
\end{eqnarray}
The explicit calculation of the one-loop diagrams in Fig.~\ref{Fig_Dip} leads to the concrete expressions for the Wilson coefficients for these effective operators in the present model~\cite{Bai:2014osa,Fitzpatrick:2010br,DelNobile:2014eta}, given by
\begin{itemize}
\item Component-Conserving Couplings:
\begin{eqnarray}\label{CC}
c_{1i}^c = \frac{-e}{64\pi^2 m_\eta^2} \sum_{\ell}R^c_{\ell i}\Big[ \frac{1}{2} + \frac{2}{3}\ln\Big(\frac{m_\ell^2}{m_\eta^2}\Big) \Big],\quad\quad c_{2i}^c = \frac{-e}{64\pi^2 m^2_\eta}\frac{1}{4}\sum_\ell R^c_{\ell i},
\end{eqnarray}
\end{itemize}
where $R^c_{\ell i} = {\rm Re}(h_{\ell i}h^*_{\ell i})$ with $R^c_{\mu 1(2)} = R^c_{\tau 2(1)} = h_{L(S)}^2$, and
\begin{itemize}
\item Transition Couplings:
\begin{eqnarray}\label{CPCT}
c_{1}^t = \frac{-2e}{64\pi^2 m_\eta^2} \sum_{\ell}R^t_{\ell}\Big[ \frac{1}{2} + \frac{2}{3}\ln\Big(\frac{m_\ell^2}{m_\eta^2}\Big) \Big],\quad\quad c_{2}^t = \frac{-2e}{64\pi^2 m^2_\eta}\frac{1}{4}\sum_\ell R^t_{\ell},
\end{eqnarray}
\end{itemize}
where $R^t_{\ell} = {\rm Re}(h_{\ell 1}h^*_{\ell 2})=h_L h_S $. Note that the formulas in Eqs.~(\ref{CC}) and (\ref{CPCT}) only differs a factor 2. Nevertheless, at the amplitude level, such a difference would be compensated when differentiating the two identical Majorana particle fields in the component-conserving effective operators ${\cal O}^c_{1i}$ and ${\cal O}^c_{2 i}$.

With the above effective operators, the two DM particles couple to the charge and magnetic dipole moment of the nucleus, which give the spin-independent and spin-dependent differential cross sections
\begin{eqnarray}\label{XSI}
\frac{d\sigma^E_T}{d E_R} = \frac{1}{2}\Big[(c^c_{11} - 2c^c_{21})^2+ (c^c_{12} - 2c^c_{22})^2 + \frac{2}{4}(c^t_1-2c^t_2)^2\Big] e^2 Z^2 \frac{m_T}{4\pi} \Big( 2-\frac{m_T E_R}{\mu_{NT}^2 v^2} \Big) F_E^2(q^2),~~
\end{eqnarray}
and
\begin{eqnarray}\label{XSD}
\frac{d\sigma^M_T}{d E_R} = \frac{1}{2}\Big[(c^c_{11} - 2c^c_{21})^2+ (c^c_{12} - 2c^c_{22})^2 + \frac{2}{4}(c^t_1-2c^t_2)^2\Big] e^2 \frac{1}{2\pi} \frac{E_R}{v^2} \frac{m_T^2 \lambda^2_T}{m_N^2 \lambda_N^2} \frac{J_T+1}{3 J_T} F_M^2(q^2),~~
\end{eqnarray}
where $\lambda_{N(T)} = e/(2 m_{N(T)})$ is the magneton of the nucleon (target nucleus), $m_{N(T)}$ is the nucleon (target nucleus) mass, $J_T$ ($Z$) is the spin (charge) of the target nucleus, and $F_{E(M)}(q^2)$ is the form factor of the nucleus charge (magnetic dipole moment), respectively. Note that in Eqs.~(\ref{XSI}) and (\ref{XSD}), the factor 1/2 accounts for the half density carried by each DM component in the Universe, while 2/4 the right normalization of the transition scattering in accord with the previous two component-conserving ones.

From Eqs.~(\ref{XSI}) and (\ref{XSD}), it is obvious that the spin-independent part is the dominant one due to the large charge $Z^2$ enhancement for the heavy target nuclei, such as Xenon used in the LUX experiments. However, the differential cross section in the recoil energy $E_R$ is suppressed by an additional power of $v^2$ or $E_R$ compared with the usual Dirac fermionic DM case, resulting in a weak direct detection constraint. An estimate made in Ref.~\cite{Bai:2014osa} for $^{129}_{54}{\rm Xe}$ with the typical reference recoiled energy and the similar masses of $\chi_{1,2}$ and $\eta$ gives the DM-nucleon scattering cross section of ${\cal O}(10^{-49}){\rm cm}^2$, which is evidently much smaller than the current LUX bound for the spin-independent DM-nucleon cross section~\cite{LUX}. Therefore, we do not show the direct DM detection constraints on our model.
% in Fig.~\ref{Fig_Result}.

Finally, we point out that if the restriction on the real Yukawa couplings is relaxed, the same one-loop Feynman diagrams in Fig.~\ref{Fig_Dip} would also induce the following transition operators:
\begin{eqnarray}\label{CPoddO}
{\cal O}^t_3 = \bar{\chi_2} \gamma^\mu \partial^\nu \chi_1 F_{\mu\nu}+ {\rm h.c.}, \quad\quad {\cal O}^t_4 = -i \bar{\chi_2}\gamma^\mu\gamma^5 \partial^\nu \chi_1 F^{\alpha\beta}\epsilon_{\mu\nu\alpha\beta} + {\rm h.c.},
\end{eqnarray}
with the corresponding Wilson coefficients
\begin{eqnarray}\label{CPVT}
c_{3}^t = \frac{-2e}{64\pi^2 m_\eta^2} \sum_{\ell}I^t_{\ell}\Big[ \frac{1}{2} + \frac{2}{3}\ln\Big(\frac{m_\ell^2}{m_\eta^2}\Big) \Big],\quad\quad c_{4}^t = \frac{-2e}{64\pi^2 m^2_\eta}\frac{1}{4}\sum_\ell I^t_{\ell},
\end{eqnarray}
where $I^t_{\ell} = {\rm Im}(h_{\ell 1}h^*_{\ell 2})$. If the imaginary part of $h_{\ell i}$ is the same order as the real one, the dominant contribution to the direct detection experiments is given by the spin-independent cross section induced by the operator ${\cal O}^{\,t}_3$. In this case, there is no suppression caused by the small recoiled energy $E_R$ or the DM velocity $v^2$, and in turn the LUX experiments have already given strong constraints to this model, effectively ruling out all the parameter space trying to explain the X-ray excess.

\subsection{Collider Searches}\label{collider}
A large portion of the parameter space of the present model has already been probed by the LEP and LHC experiments. Since our two DM components do not directly couple to the electron at tree-level by the construction, they cannot be directly pair produced at the LEP so that the relevant monophoton constraints~\cite{Fox:2011fx} do not apply.

Another collider signal in our model is related to the production of the singly-charged scalar $\eta$ mediated by a virtual photon or $Z$-boson. When the mass of $\eta$ is lighter than about 100 GeV, it can be pair produced at the $e^+ e^-$ colliders, which has been already ruled out by the null results at the LEP. Since the maximum energy of the LEP is about 200 GeV, we take 100 GeV as the LEP bound for the charged mediator mass as an illustration in our study. At the LHC, the signal of this leptophilic model is the creation of the charged mediator pair via the Drell-Yan process, each of which decays further into a lepton plus one of the DM particles~\cite{Bai:2014osa,Agrawal:2014ufa,Chang:2014tea}. The signature at hadron colliders is a pair of opposite-sign leptons with some missing transverse energy. Such a signature is very similar to that of searching for the right-handed sleptons with mixings in the MSSM at the colliders, for which the current LHC bound of the mass of the charged scalar is about 240 GeV as presented in Fig.~8a in Ref.~\cite{ATLAS}, which is more stringent than the LEP direct search bound.

In this work, we use both LEP and LHC bounds in our numerical studies, which can be transformed into the mass bounds of the two nearly degenerate DM components for a given mass ratio $R_M= m_\eta/m_\chi$. In Fig.~\ref{Fig_Result}, the excluded regions for both experimental bounds are plotted as the purple and yellow regions bounded by the dot-dashed and dotted lines. Note that the latter LHC constraint can be relaxed a lot for the present model. The LHC bound of 240 GeV for the charged scalar $\eta$ is the most stringent one in Fig.~8a in Ref.~\cite{ATLAS}. The actual restricted region in the $m_\eta$-$m_\chi$ plane is much irregular, and the constraint becomes more loose when the DM mass approaches the mediator one due to the decreasing of the missing energy. In Fig.~\ref{Fig_ATLAS}, we show the actual LHC bound on the $m_\eta$-$m_\chi$ plane given in Ref.~\cite{ATLAS}, together with the allowed X-ray signal region and other constraints with the Yukawa couplings $(h_S, h_L)=(0.113,1.13)$. Another important issue is the assumption made in the analysis of Ref.~\cite{ATLAS} that the decay of the charged scalar is only to the first two generations of the charged leptons plus a large missing energy. Nevertheless, in our model, $\eta$ can only have about one fourth probability to $\mu^+ + \mu^-$, whereas the other 3/4 of the decay branching ratio related to the $\tau$ lepton cannot be probed. As the detailed analysis of the LHC bound on our model is beyond the scope of the present paper, we are content with the most stringent bound in Ref.~\cite{ATLAS} in our exploration of the model parameter space.

\section{Final Results}\label{result}
\begin{figure}
\centering
\includegraphics[width=1\textwidth, natwidth=1170,natheight=700]{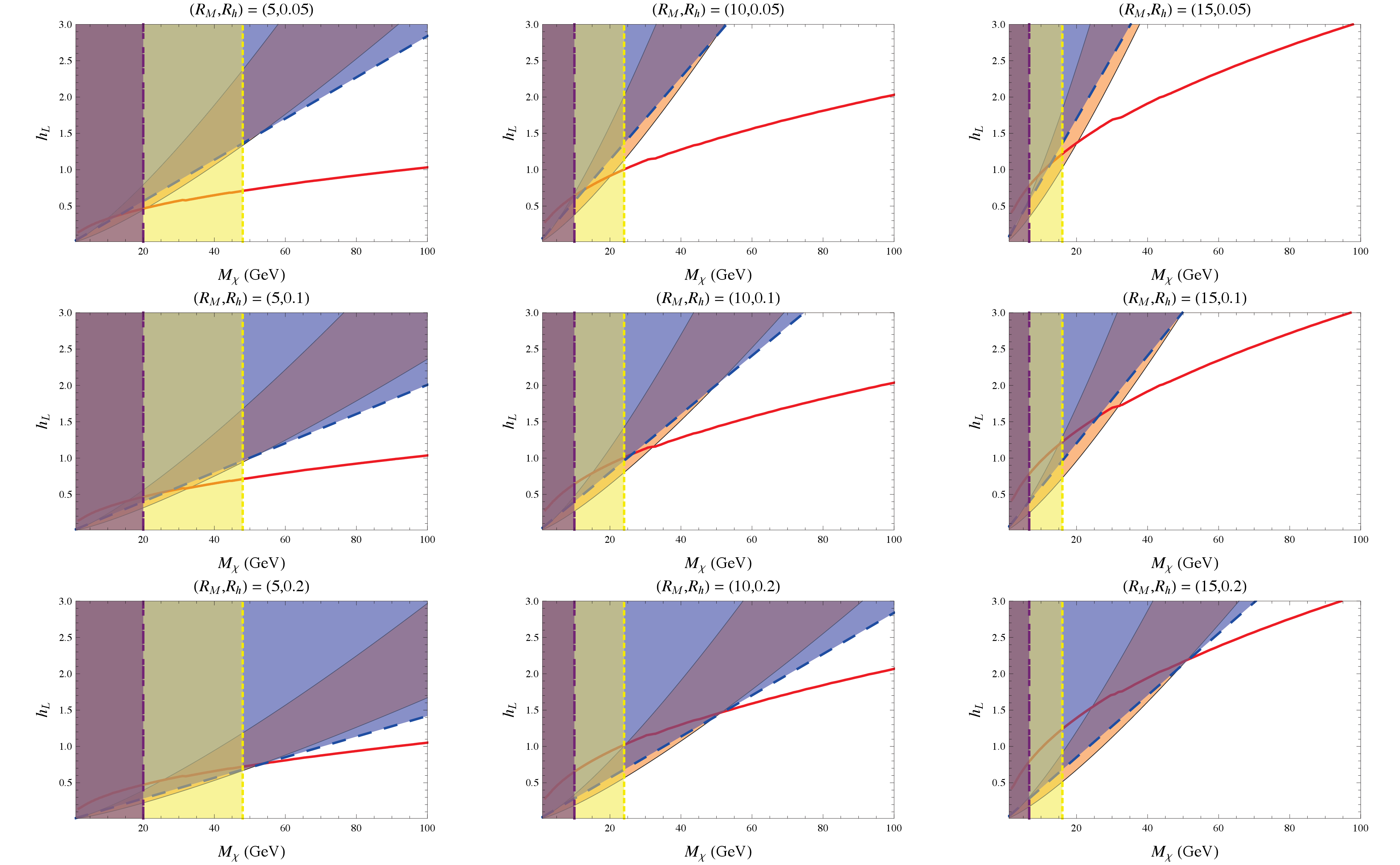}
\caption{Parameter spaces in the $m_\chi$-$h_L$ plane for $R_M=m_\eta/m_N=(5,\,10,\,15)$ and $R_h=h_S/h_L= (0.05,\,0.1,\,0.2)$, where the thick red curve gives the correct DM relic abundance, the orange region bounded by the thin black curves is accord with the X-ray line observation, and the blue, yellow and purple regions bounded by the dashed, dotted and dot-dashed lines are excluded by the FLV process $\tau\to\mu\gamma$, LEP and LHC, respectively.
}\label{Fig_Result}
\end{figure}

Our results on the parameter space are shown in Fig.~\ref{Fig_Result} for different choices of the mass ratios $R_M=m_\eta/m_\chi$ and the Yukawa coupling ratios $R_h=h_S/h_L$ . The orange region with the thin black boundaries represents the parameter space which can be used to explain the X-ray line fluxes, while the thick red curve gives the correct DM relic abundance. The blue, yellow and purple regions bounded by the dashed, dotted and dot-dashed lines are excluded by the constraints from the $\tau\to\mu\gamma$ process, LEP and LHC, respectively. A further constraint is originated from the perturbativity requirement of the large Yukawa coupling $h_L$, which is assumed to be less than 3.

\begin{figure}[th]
\centering
\includegraphics[width=0.4\textwidth,natwidth=300,natheight=300]{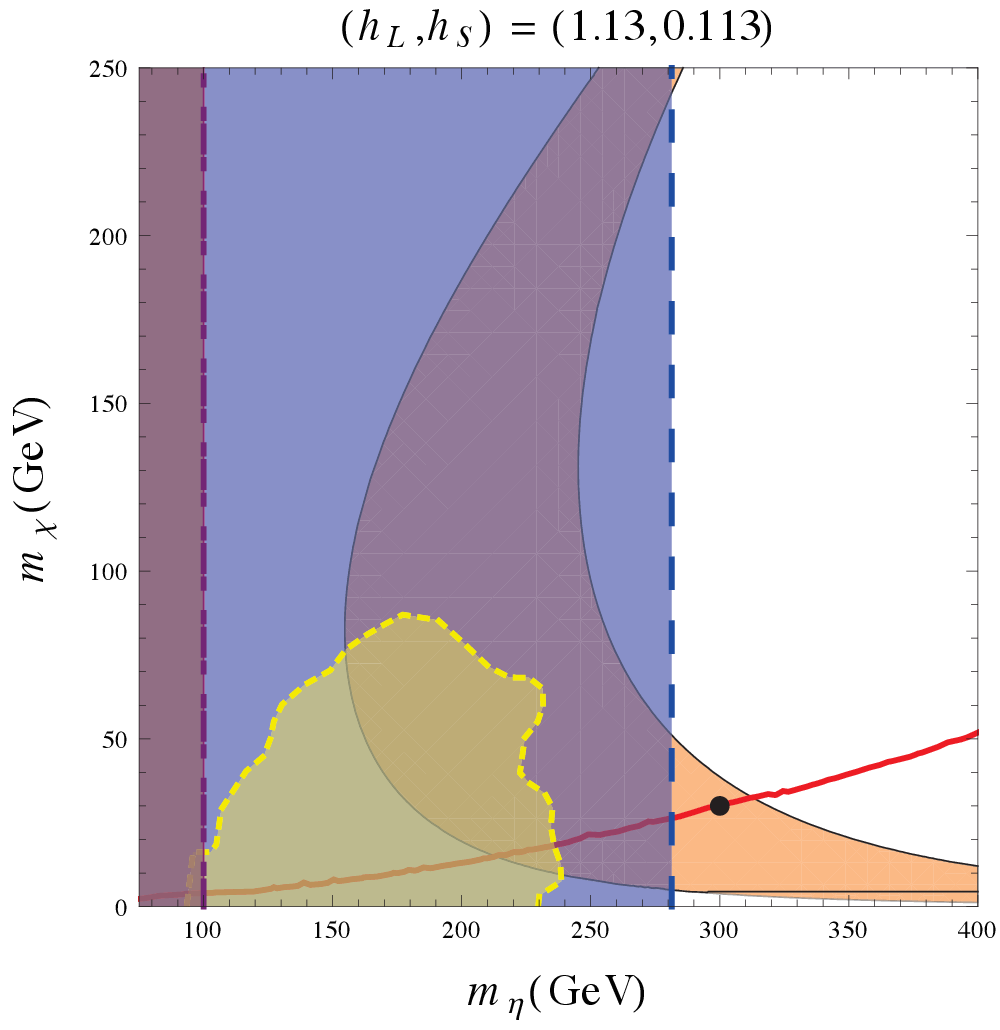}
\caption{Label is the same as Fig.~\ref{Fig_Result} but in the $m_\eta$-$m_\chi$ plane for ($h_L$,$h_S$)=(1.13,0.113). The black point represents the benchmark point given in the context.}
\label{Fig_ATLAS}
\end{figure}

In Fig.~\ref{Fig_Result}, we arrange the diagrams in rows and columns in terms of the mass ratios $R_M$ and the coupling ratios $R_h$ in order to show the trend of how the allowed parameter space changes when these two parameters vary. It is seen that there are still parameter spaces in which the present two-component DM model can explain the cluster X-ray line and DM relic abundance in the Universe, simultaneously, despite of the very strong constraints from flavor and collider physics. As an illustration, we present a prototypical benchmark point: $h_S=0.113$, $h_L=1.13$, $m_\chi=30~$GeV and $m_\eta = 300~$GeV. Fig.~\ref{Fig_ATLAS} illustrates the parameter space in the $m_\eta$-$m_\chi$ plane with the fixed Yukawa couplings ($h_L,~h_S$)=(1.13,0.113), in which the current ATLAS bounds on $m_\eta$ and $m_\chi$ are shown more clearly.

\section{Conclusions and Discussions}\label{conclusion}
We have proposed a two-component DM model to understand the recent unidentified X-ray line from the observation of the Andromeda galaxy and the distant galaxy clusters. In order for the two DM particles $\chi_{1,2}$ to only couple to the right-handed $\mu$ and $\tau$ via the Yukawa couplings, a new singly-charged scalar mediator $\eta$ is introduced. Both $\chi_{1,2}$ and $\eta$ are odd under the dark $Z_2$ symmetry. In spite of the very stringent constraints from flavor physics, direct DM searches and collider bounds from the LEP and LHC, we have found some parameter space which can explain the X-ray line with the correct DM relic abundance.

%{\color{red} 
As seen from Eq.~(5), the successful explanation of the cluster X-ray flux depends on the 3.5 keV mass splitting $\Delta m$ between the two DM components, which is naturally motivated by the measured energy of the X-ray line. The next question is to investigate the stability of such a small splitting against the radiative corrections, especially when we notice the DM Majorana masses are of ${\cal O}$(10 GeV). However, such a worry can be avoided for the present DM model. Since the leading order radiative correction to the DM Majorana masses appears at three-loop level, we find that the correction to the mass splitting is always of order sub-keV, which is achieved by the suppressions from the loop factors and the various couplings. Therefore, we expect that the keV order of the DM mass splitting is naturally guaranteed against the radiative corrections in our model.
%}

In our study, we have used the most stringent bound from the ATLAS~\cite{ATLAS} to constrain the charged scalar mediator mass $m_\eta$. However, this constraint can be relaxed if we take into account the difference of the present model from the assumptions made in Ref.~\cite{ATLAS}. This analysis is helpful because this constraint excludes all or part of the signal region in some parameter spaces, such as those shown in the second and third graphs in the first row of Fig.~\ref{Fig_Result}. A detailed investigation of this collider constraint may reopen these parameter windows, which will be worth of a future work.

\section*{Acknowledgments}
The work was supported in part by National Center for Theoretical Sciences, National Science
Council (NSC-101-2112-M-007-006-MY3) and National Tsing Hua
University (Grant Nos. 103N1087E1 and 103N2724E1).

\end{document}